# Microsecond-resolved electro-optic dual-comb spectroscopy in the 10–12.5 µm fingerprint region for radical kinetics

PEI-LING LUO[1,*] AND I-YUN CHEN[1,2]

[1] *Institute of Atomic and Molecular Sciences, Academia Sinica, Taipei 106319, Taiwan.*
[2] *Department of Chemistry, National Taiwan University, Taipei, 10617, Taiwan.*
**plluo@gate.sinica.edu.tw*

**Abstract:** Dual-comb spectroscopy enables broadband, high-resolution measurements with microsecond temporal resolution, but extending this capability to the 10–12.5 µm molecular fingerprint region remains technically challenging, particularly for transient radical kinetics. Here, we demonstrate microsecond-resolved dual-comb spectroscopy in this spectral range using electro-optic combs and difference-frequency generation in an orientation-patterned gallium phosphide crystal. Operation near a turning-point quasi-phase-matching condition at approximately 140 °C reduces the wavelength sensitivity of the nonlinear conversion, enabling robust tuning of the idler comb over 83 cm$^{-1}$, corresponding to approximately 1.2 µm near 12 µm, by adjusting only the signal-comb center wavelength while keeping the pump wavelength and crystal temperature fixed. As a demonstration, we perform high-resolution, microsecond-resolved spectroscopy of transient chlorine monoxide (ClO) near 12 µm. Time-resolved dual-comb spectra capture the temporal evolution of ClO produced by the Cl + O$_3$ reaction with a temporal resolution of 1.5 µs, enabling quantitative determination of the ClO formation rate coefficient. These results establish this dual-comb platform as a promising tool for quantitative, microsecond-resolved studies of short-lived radicals, particularly atmospherically relevant halogen oxides.

## 1. INTRODUCTION

Dual-comb spectroscopy combines broadband spectral coverage, high spectral resolution, and rapid acquisition, making it a powerful tool for probing transient chemical processes in gas-phase [1–6], condensed-phase [7,8], and plasma systems [9,10]. In the mid-infrared, roughly spanning 3–15 µm, these capabilities are especially valuable because this spectral region probes fundamental molecular vibrations, enabling high sensitivity and selectivity in species-resolved measurements. In particular, the short-wave mid-infrared (3–5 µm) is well suited for C–H, O–H, and N–H stretching vibrations, whereas the long-wave fingerprint region (6–15 µm) provides access to bending and deformation modes that are often more diagnostic of molecular structure.

Several complementary approaches toward time-resolved mid-infrared dual-comb spectroscopy have emerged in recent years [1–9]. In the short-wave mid-infrared, electro-optic (EO) dual-comb systems combined with difference-frequency generation (DFG) [3,4] or optical parametric oscillators (OPOs) [1] have enabled ultrafast, high-resolution measurements with agile wavelength tuning across 2.2–4.6 µm. For example, time-resolved dual-comb absorption spectroscopy with a comb-line spacing of 2.55 GHz has been used to monitor a $CO_2$ absorption transition in a supersonic pulsed jet with 20 ns temporal resolution and a single-acquisition spectral span of approximately 1 cm$^{-1}$ [1]. In addition, dual-comb spectrometers based on 1-GHz mode-locked frequency combs with DFG have enabled broadband studies of fast chemical kinetics across the 3–5 µm region, with continuous optical coverage exceeding 1000 cm$^{-1}$ and acquisition times as short as 17.5 µs for a 290 cm$^{-1}$ spectral span per detector, as demonstrated in shock-tube studies of trioxane ($C_3H_6O_3$) decomposition by monitoring $C_3H_6O_3$ and formaldehyde (HCHO) near 3.5 µm and carbon monoxide (CO) near 4.5 µm [2]. These measurements also indicate a fundamental trade-off among spectral bandwidth, spectral resolution, and time resolution in time-resolved dual-comb spectroscopy, in which the shortest acquisition times are achieved by reducing the instantaneous spectral span per detector through spectral filtering or parallel detection.

In the long-wave mid-infrared, quantum cascade laser (QCL)-based dual-comb spectroscopy has been used for studies of combustion diagnostics [6], surface electrochemistry [7], and protein dynamics [8], although many reported time-resolved demonstrations have been carried out in relatively narrow spectral windows centered near 8 µm. These compact platforms benefit from multi-GHz comb-line spacing, high optical power, and sub-microsecond to microsecond time resolution [11]. Such characteristics are particularly advantageous for pressure-broadened or condensed-phase measurements, where broad spectral features of the sample reduce the demand for dense spectral sampling, whereas the large comb-line spacing and limited per-device spectral bandwidth make broadband, high-resolution gas-phase spectroscopy of narrow transitions less straightforward without spectral interleaving or tuning. In parallel, DFG-based electro-optic dual-comb spectroscopy has been used to probe short-lived Criegee intermediates near 7.77–8.22 µm, demonstrating that near-infrared EO combs combined with DFG are well suited for high-resolution, time-resolved measurements in the long-wave mid-infrared [5]. However, extension to longer wavelengths in DFG-based dual-comb implementations is typically constrained by the conversion efficiency and phase-matching bandwidth of the nonlinear frequency-conversion stage. Separately, broadband mid-infrared comb sources with spectral coverage extending well beyond 8 µm have been developed based on intra-pulse difference-frequency generation (IP-DFG) in nonlinear crystals [12–14]. For example, octave-spanning mid-infrared comb generation over 4–12 µm has been demonstrated using few-cycle laser pulses combined with IP-DFG [12], while long-wave dual-comb platforms have enabled comb-line-resolved spectroscopy over 6.6–11.4 µm for high-precision measurements of stable molecules such as $N_2O$ and $CH_3OH$ [13]. Despite this substantial spectral progress, quantitative, microsecond-resolved dual-comb measurements of transient species beyond 10 µm have not yet been achieved.

In this work, we implement microsecond-resolved dual-comb spectroscopy in the 10–12.5 µm molecular fingerprint region using electro-optic combs with widely tunable center wavelengths in the telecom spectral range, combined with difference-frequency generation in an orientation-patterned gallium phosphide (OP-GaP) crystal. Operation near a turning-point quasi-phase-matching condition enables stable and continuous spectral tuning over a wide range near 12 µm. We further apply this dual-comb platform to study chlorine monoxide (ClO), a key reactive radical in terrestrial ozone chemistry and a species of interest in Venusian atmospheric chemistry [15–18]. To date, atmospheric measurements of ClO have been carried out using microwave limb sounding [19] and infrared limb-emission Fourier-transform spectroscopy [20,21], whereas laboratory kinetic studies have employed a range of methods, notably time-resolved ultraviolet absorption [22,23] and mass-spectrometric techniques [24]. Infrared studies of ClO, by contrast, have mainly relied on Fourier-transform infrared spectroscopy (FTIR) for spectral characterization [25–27], while quantitative infrared measurements of its rapid kinetics have remained largely unexplored. Here, we investigate ClO formation in the Cl + $O_3$ reaction system through time-resolved monitoring of its rovibrational transitions near 12 µm, demonstrating the capability of this dual-comb platform for quantitative laboratory kinetic studies of halogen-oxide chemistry.

## 2. RESULTS

### A. Electro-optic dual-comb source in the 10–12.5 µm molecular fingerprint region

Figure 1 summarizes the experimental setup and representative spectral characteristics of the electro-optic (EO) dual-comb platform in the 10–12.5 µm molecular fingerprint region. The schematic of the system is shown in Fig. 1(a). The idler dual-comb source is generated by quasi-phase-matched (QPM) difference-frequency generation (DFG) between a 1380 nm continuous-wave (cw) pump beam and a tunable telecom-band EO dual-comb signal beam. The 1380 nm pump laser is power-amplified to 5 W using a Raman fiber amplifier (RFA). The telecom-band EO dual-comb is generated from a mode-hop-free tunable external-cavity diode laser (Santec, TSL-570-H) using electro-optic intensity modulators (Optilab, IML-1550-40-PM-V-HER) driven by 25 ps electrical pulse generators (Alnair Labs, EPG-220B). The repetition frequency ($f_{rep}$) of the pulse generators determines the comb-line spacing and can be tuned from 0.1 to 1 GHz. After amplification to 2 W, the EO dual-comb signal beam is combined with the pump beam and focused into the OP-GaP crystal for nonlinear down-conversion into the long-wave mid-infrared near 12 µm.

Figure 1(b) shows a representative comb-line-resolved dual-comb spectrum near 860.6 cm$^{-1}$, obtained by Fourier transformation of idler dual-comb interferograms detected with a HgCdTe detector (MCT; Vigo, PVI-4TE-10.6) and digitized using a 12-bit data-acquisition board (DAQ; AlazarTech, ATS9371) at 500 MS/s. The idler dual-comb source covers more than 1.5 cm$^{-1}$, corresponding to 45 GHz, with an average optical power of approximately 50 µW. With a comb-line spacing ($f_{rep}$) of 210 MHz and a repetition-rate difference (Δf) of 0.32 MHz, more than 160 high-SNR comb lines spanning approximately 1.1 cm$^{-1}$ are clearly resolved. The peak comb-line intensities exceed the background level by more than 20 dB, demonstrating sufficient spectral quality for molecular absorption measurements.

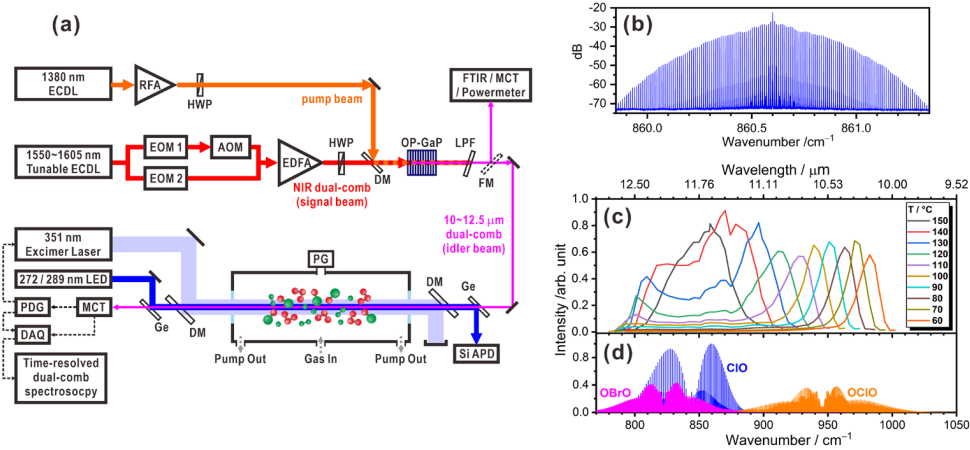

Fig. 1. (a) Schematic of the experimental setup. ECLD, external-cavity diode laser; RFA, Raman fiber amplifier; HWP, half-wave plate; EOM, electro-optic modulator; AOM, acousto-optic modulator; EDFA, erbium-doped fiber amplifier; OP-GaP, orientation-patterned gallium phosphide; LPF, long-pass filter; FM, flip mirror; FTIR, Fourier-transform infrared spectrometer; MCT, HgCdTe detector; Ge, germanium window; DM, dichroic mirror; APD, avalanche photodiode; PG, pressure gauge; LED, light-emitting diode; PDG, pulse delay generator; DAQ, data acquisition board. (b) Representative comb-line-resolved dual-comb spectrum near 860.6 cm$^{-1}$. The dual-comb spectrometer is operated with a comb-line spacing ($f_{rep}$) of 210 MHz and a repetition rate difference ($\Delta f$) of 0.32 MHz. (c) Continuously tunable idler-comb spectra obtained by tuning the signal-comb center wavelength for different OP-GaP crystal temperatures while keeping the pump wavelength fixed. (d) Representative absorption features of selected halogen oxides in the 10–12.5 μm fingerprint region, where the ClO, OClO, and OBrO spectra are simulated using PGOPHER [28]; the molecular parameters for ClO and OClO are taken from the literature [25–27,29,30], whereas those for OBrO are calculated at the B3LYP/Def2TZVPP level with Gaussian 16 [31].

Figures 1(c) and 1(d) illustrate the tunable idler-comb range achieved in the present system and the spectroscopic relevance of this long-wave window for studies of halogen oxides. The overall tunable idler-comb range is obtained by varying the signal-comb center wavelength for OP-GaP crystal temperatures between 60 and 150 °C while keeping the pump wavelength fixed, as shown in Fig. 1(c). As the crystal temperature increases, the intensity maximum of the phase-matched idler comb, denoted pk1, shifts toward longer wavelengths. At approximately 110 °C, a second phase-matched feature, pk2, appears on the longer-wavelength side, separated from pk1 by approximately 130 cm$^{-1}$. As the crystal temperature is further increased from 110 to 150 °C, the spectral separation between pk1 and pk2 gradually decreases. At 140 °C, the tunable idler-comb bandwidth near 12 μm reaches a full width at half maximum of 83 cm$^{-1}$, corresponding to approximately 1.2 μm. This behavior suggests operation near a turning-point quasi-phase-matching condition, as analyzed in Fig. 2.

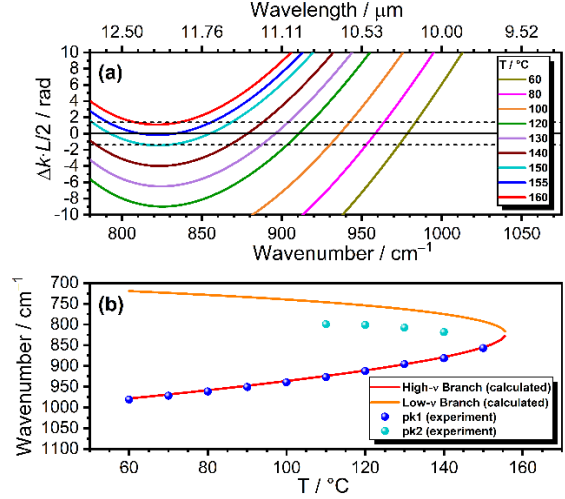

Fig. 2. Turning-point quasi-phase matching in OP-GaP near 12 μm. (a) Calculated phase-mismatch curves, $\Delta k \cdot L/2$, versus idler wavenumber for different OP-GaP temperatures. The solid horizontal line ($\Delta k \cdot L/2 = 0$) represents exact phase matching, and the dashed lines ($\Delta k \cdot L/2 = \pm 1.39$) define the phase-matching acceptance bandwidth. Near 140–150 °C, the phase-mismatch curves exhibit a turning point, resulting in reduced wavelength sensitivity. (b) Calculated and measured temperature-dependent phase-matched idler wavenumbers. The red and orange curves correspond to the calculated high-ν and low-ν branches, respectively, while blue and cyan symbols represent the peak positions (pk1 and pk2) extracted from the tunable idler-comb spectra in Fig. 1(c).

Figure 2(a) shows calculated phase-mismatch curves, expressed as $\Delta kL/2$, as a function of idler wavenumber for different OP-GaP crystal temperatures. The phase mismatch is defined as $\Delta k = k_p - k_s - k_i - 2\pi/\Lambda(T)$, where $k_p$, $k_s$, and $k_i$ are the wave vectors of the pump, signal, and idler fields, respectively. The length of the OP-GaP crystal, $L$, is 20 mm. The temperature-dependent grating period is given by $\Lambda(T) = \Lambda_{25°C}[1+\alpha(T-25\ °C)]$, with $\Lambda_{25°C} = 52.5$ μm and $\alpha = 5\times10^{-6}$ K$^{-1}$. These curves are calculated using the quasi-phase-matching relation together with the temperature-dependent Sellmeier equation for GaP given in Eq. (3), with the coefficients listed in Table 2 of Ref. [32]. The solid horizontal line at $\Delta kL/2 = 0$ indicates exact phase matching, while the dashed lines at $\Delta kL/2 = \pm 1.39$ define the phase-matching acceptance bandwidth. Within this acceptance band, the broadest usable idler-wavenumber range is obtained near 140–150 °C, indicating that the phase-matched idler wavelength becomes less sensitive to changes in the tuning parameters. This behavior broadens the usable QPM window and accounts for the large idler-comb tuning bandwidth observed near 12 μm.

Figure 2(b) compares the calculated phase-matched idler wavenumbers with the experimentally observed peak positions extracted from Fig. 1(c). The primary peak, pk1, follows the calculated high-ν branch closely, confirming that the main phase-matched output is well described by the QPM model. By contrast, the secondary peak, pk2, shows a systematic offset from the second $\Delta k = 0$ solution calculated under the assumption of a uniform crystal temperature. A possible interpretation is that pk2 reflects phase-matching contributions from crystal regions at slightly higher effective temperatures than the nominal set temperature, suggesting temperature nonuniformity along the crystal. This qualitative picture may also be influenced by the increasing absorption of OP-GaP toward the long-wavelength edge of the accessible range, where the absorption coefficient rises to approximately 2 cm$^{-1}$ near 12.5 μm [33]. As a result, the observed pk2 feature may not directly trace the ideal phase-matching condition of the low-ν branch.

By operating near the turning-point QPM condition with a single-period OP-GaP crystal over 60–150 °C, the present system achieves an accessible idler range exceeding 200 cm$^{-1}$ near 12 μm. For comparison, a previously reported DFG-based electro-optic dual-comb system near 8 μm, using 1590–1610 nm tunable combs mixed with a 1999 nm cw laser, achieved an idler range of approximately 42 cm$^{-1}$ under a comparable single-period-crystal tuning scheme [5]. Although the two systems operate in different wavelength regions, the substantially broader range achieved here highlights the effectiveness of the turning-point QPM condition for expanding accessible tuning in the longer-wavelength fingerprint region. This broad tunability directly addresses a key phase-matching bandwidth limitation for broadband DFG-based dual-comb measurements in the 10–12.5 μm fingerprint region. It also enables spectroscopic access to several halogen oxides, including ClO, OClO, and OBrO, as shown in Fig. 1(d).

## B. High-resolution, time-resolved dual-comb spectroscopy of ClO near 12 μm

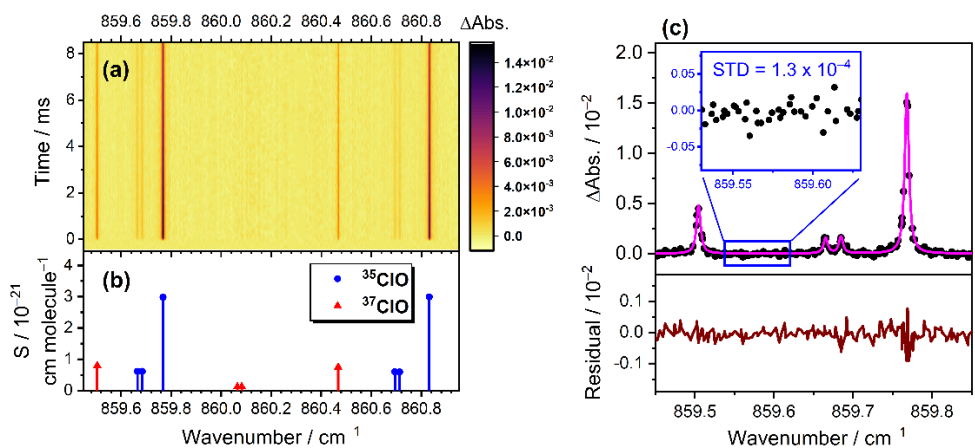

Fig. 3. (a) Representative time-resolved dual-comb spectra of ClO in the 859.45–860.95 cm$^{-1}$ region following 351-nm irradiation of a flowing mixture of Cl$_2$/O$_3$/O$_2$/N$_2$ (2.4/0.2/25.9/5.0 Torr, $P_T$ = 33.6 Torr, 296 K). The spectra are recorded with a temporal resolution of 125 μs and an average spectral resolution of 0.002 cm$^{-1}$ by interleaving eight dual-comb spectra acquired at different center wavelengths. Each measurement is averaged over 10000 excimer laser shots. The dual-comb spectrometer is operated with a comb-line spacing ($f_{rep}$) of 210 MHz and a repetition rate difference (Δf) of 0.32 MHz. (b) Stick spectra of ClO from the literatures [25–27,34]. (c) Rotationally resolved ClO spectrum recorded 125–250 μs after photolysis. The pink and brown curves show the multi-peak fit and fit residual, respectively. The ClO concentration derived from the fitted integrated absorbance areas and corresponding line strengths is 4.6×10$^{14}$ molecule cm$^{-3}$. The detection limit is estimated to be 3.7×10$^{12}$ molecule cm$^{-3}$ for the absorption peak at 859.768 cm$^{-1}$ over a 58-cm absorption path.

Chlorine monoxide (ClO) is a linear radical with fundamental Cl–O stretching bands centered at 835.468 and 842.565 cm$^{-1}$ for $^{35}$ClO and $^{37}$ClO, respectively [29]. In laboratory studies, ClO radicals can be generated by several methods, including reactions of Cl atoms with O$_3$ or ClO$_2$ [25–27]. Although ClO is a transient radical, its self-reaction loss rate can be much smaller than its formation rate under low-pressure, high-flow conditions. By operating the reaction cell in a regime where the gas residence time is comparable to the ClO lifetime, its rovibrational features can be measured under quasi-steady-state conditions using high-resolution FTIR spectroscopy. Indeed, high-resolution spectra of ClO in the $X^2\Pi$–$X^2\Pi$ (1–0) band have previously been measured with a spectral resolution of 0.004 cm$^{-1}$ to derive accurate rovibrational constants [27], and absolute line strengths have been reported with an uncertainty of approximately 15% [25]. However, conventional FTIR spectroscopy is less suited for direct measurements of rapid reaction kinetics, such as determining the rate coefficient for ClO formation via the Cl + O$_3$ reaction. Here, we combine the present 12 μm dual-comb source with a flash-photolysis reaction cell to perform high-resolution, time-resolved spectroscopy of ClO.

The reaction cell consists of a glass reactor equipped with KBr windows on both sides, allowing simultaneous coupling of the 12 μm dual-comb beam, the 351 nm photolysis beam, and ultraviolet light-emitting diodes (LEDs) used for in situ $O_3$ measurements, as shown in Fig. 1(a). ClO is generated by 351 nm excimer-laser photolysis of flowing $Cl_2/O_3/O_2/N_2$ mixtures. Upon photolysis, $Cl_2$ produces Cl atoms, which subsequently react with $O_3$ to form ClO and $O_2$. $O_3$ is generated by passing mixed $O_2$ and $N_2$ flows, regulated by calibrated mass flow controllers (MKS 1179A), through a quartz discharge cell. The $O_3$ mixing ratio is controlled by the discharge current and the $O_2/N_2$ ratio and quantified from FTIR spectra recorded in the 950–1100 $cm^{-1}$ region with a 10 cm cell before the $O_3$-containing flow is mixed with the main $Cl_2$ and $O_2$ flows upstream of the reactor. Small $N_2$ purge flows are used to protect the reactor windows and are exhausted without mixing with the main flow. The effective absorption path length of 58 cm is calibrated with an uncertainty of less than 3% using a standard reference gas such as $NH_3$. Precursor concentrations in the reactor are determined from the flow rates, premix ratios, and measured total pressure.

Figures 3(a) and 3(b) show representative time-resolved dual-comb spectra of ClO in the 859.45–860.95 $cm^{-1}$ region together with the corresponding stick spectra of $^{35}$ClO and $^{37}$ClO. The time-resolved dual-comb spectra are recorded with a temporal resolution of 125 μs using the present dual-comb platform with a comb-line spacing ($f_{rep}$) of 210 MHz and a repetition rate difference ($\Delta f$) of 0.32 MHz. An average spectral resolution of 0.002 $cm^{-1}$ is achieved by interleaving eight dual-comb spectra acquired at different center wavelengths. Several absorption features of both $^{35}$ClO and $^{37}$ClO are clearly identified, including $^{37}$ClO $X^2\Pi_{3/2}$ R(19.5)$_{e,f}$ at 859.504 $cm^{-1}$, $^{35}$ClO $X^2\Pi_{1/2}$ R(15.5)$_{e,f}$ at 859.665 and 859.684 $cm^{-1}$, $^{35}$ClO $X^2\Pi_{3/2}$ R(12.5)$_{e,f}$ at 859.768 $cm^{-1}$, $^{37}$ClO $X^2\Pi_{3/2}$ R(20.5)$_{e,f}$ at 860.468 $cm^{-1}$, $^{35}$ClO $X^2\Pi_{1/2}$ R(16.5)$_{e,f}$ at 860.694 and 860.714 $cm^{-1}$, and $^{35}$ClO $X^2\Pi_{3/2}$ R(13.5)$_{e,f}$ at 860.832 $cm^{-1}$.

Figure 3(c) displays high-resolution differential absorbance (ΔAbs.) spectra of ClO in the 859.45–859.85 $cm^{-1}$ region, extracted from Fig. 3(a) over the time interval of 125–250 μs after 351 nm laser photolysis. The spectra are fitted using a multi-peak Voigt function to derive the integrated absorbance of each transition. By applying Beer's law, a ClO concentration of $4.6\times10^{14}$ molecule $cm^{-3}$ is derived from the fitted integrated absorbance areas and the corresponding line strengths [25–27,34]. A spectral noise floor of $1.3\times10^{-4}$ is estimated from the standard deviation (STD) of the spectral region without ClO absorption, as indicated in Fig. 3(c). The detection limit is thus estimated to be $3.7\times10^{12}$ molecules $cm^{-3}$ for the absorption peak at 859.768 $cm^{-1}$ over a 58 cm absorption path. These results demonstrate that the present dual-comb platform is capable of quantitative, high-resolution, time-resolved spectroscopy of transient ClO near 12 μm, providing the basis for subsequent kinetic measurements.

## C. Kinetic measurements of ClO formation in the Cl + $O_3$ reaction system

To further evaluate the capability of the present dual-comb platform for kinetic measurements, the temporal evolution of the ClO absorption line near 859.768 $cm^{-1}$ is monitored following 351 nm photolysis of flowing $Cl_2/O_3/O_2/N_2$ mixtures at different initial $O_3$ concentrations. For microsecond-resolved measurements, a larger comb-line spacing ($f_{rep}$) is used to allow a higher repetition-rate difference ($\Delta f$) and thus improve the temporal resolution. In the present EO dual-comb system, using $f_{rep}$ = 480 MHz allows $\Delta f$ to exceed 2 MHz, corresponding to a minimum temporal resolution of 0.5 μs. The larger comb spacing also increases the intensity of individual comb lines in the comb-line-resolved spectrum by approximately 6–7 dB, leading to improved spectral signal-to-noise ratio (SNR) for a given averaging time.

In time-resolved dual-comb spectroscopy, the temporal resolution can be adjusted by changing the length of the dual-comb interferogram used to generate each time-dependent Fourier spectrum. At higher temporal resolution, the measurement time associated with each spectrum becomes shorter, and the SNR of the resulting temporal profile is consequently reduced. Figure 4(a) shows the temporal profiles of the ClO differential absorbance signal at 859.768 cm$^{-1}$ extracted from time-resolved dual-comb spectra measured at a total pressure of 33.2 Torr and 296 K, with an initial O$_3$ concentration of 7.2×10$^{15}$ molecule cm$^{-3}$. After laser photolysis, ClO is observed to form rapidly via the Cl + O$_3$ reaction, reaching its maximum within less than 100 μs, and then decaying more slowly through self-reaction and other secondary processes. By analyzing the rovibrational spectra of ClO, the peak ClO concentration is estimated to be 3.9×10$^{14}$ molecule cm$^{-3}$. SNRs of 52 and 166 are obtained from the measured ClO temporal profiles averaged over 50000 photolysis laser shots at temporal resolutions of 1.5 and 15 μs, respectively. Figure 4(b) shows the evolution of the SNR and noise floor of the ClO differential absorbance temporal profile with a temporal resolution of 1.5 μs as a function of the number of spectral averages. Over 50000 averages, the noise floor of the temporal absorbance profile is reduced to 2.5×10$^{-4}$, and the corresponding detection limit for ClO is estimated to be 7.5×10$^{12}$ molecule cm$^{-3}$. Such sensitivity and microsecond temporal resolution enable further kinetic studies of ClO formation in the Cl + O$_3$ reaction system under pseudo-first-order conditions, for which the initial O$_3$ concentration is set in large excess over the Cl concentration.

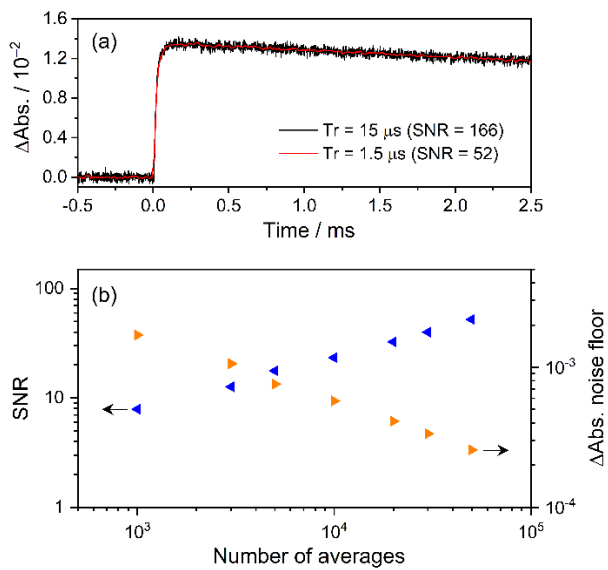

Fig. 4. (a) Comparison of the temporal profiles of ClO at 859.768 cm$^{-1}$ extracted from time-resolved dual-comb spectra averaged over 50,000 photolysis shots, with temporal resolutions of 1.5 μs (black) and 15 μs (red). The dual-comb spectrometer is operated with a comb-line spacing ($f_{rep}$) of 480 MHz and a repetition rate difference (Δf) of 2.0 MHz. (b) Signal-to-noise ratio (SNR) and noise floor of the differential absorbance temporal profile of the ClO absorption peak at 859.768 cm$^{-1}$ with 1.5 μs temporal resolution as a function of the number of averages. The measurements are performed at a total pressure of 33.2 Torr and 296 K, with the peak ClO concentration estimated to be 3.9×10$^{14}$ molecule cm$^{-3}$. The detection limit for monitoring the ClO temporal profile at 859.768 cm$^{-1}$ is 7.5×10$^{12}$ molecule cm$^{-3}$ at 1.5 μs temporal resolution with 50000 averages.

Figure 5(a) shows ClO temporal profiles recorded under experimental conditions with an initial Cl atom concentration of approximately 4 × 10$^{14}$ molecule cm$^{-3}$ and initial O$_3$ concentrations of (7.2–20.3) × 10$^{15}$ molecule cm$^{-3}$ at a total pressure of $P_T$ = 33.2 Torr and 296 K. Under these conditions, the subsequent decay of ClO is observed to be more than three orders of magnitude slower than its initial formation. Therefore, the early-time region of the ClO temporal profile (typically within the first 80 μs after photolysis) can be analyzed to determine the ClO formation rate coefficient. Under pseudo-first-order conditions, where $O_3$ is present in large excess over Cl atoms ($[O_3] \gg [Cl]$), the rise of the ClO signal can be described by a single-exponential function, $[ClO]_{obs}(t) = A \times [1- \exp(-k_{obs} \times t)]$, in which $[ClO]_{obs}(t)$ represents the observed time-dependent ClO signal, $A$ is the amplitude parameter, $k_{obs}$ is the fitted first-order formation rate coefficient. Figure 5(b) shows the fitted $k_{obs}$ values as a function of the initial $O_3$ concentration. A linear fit yields a slope of $(1.04 \pm 0.02) \times 10^{-11}$ cm$^3$ molecule$^{-1}$ s$^{-1}$, corresponding to the bimolecular rate coefficient for the Cl + $O_3$ reaction, i.e., the ClO formation rate coefficient. Considering the fitting uncertainty of the slope (2%), together with additional uncertainties in the $O_3$ concentration measurement (4%), absorption path length (3%), and temperature (1%), the overall standard error is estimated to be ~6%. Accordingly, the ClO formation rate coefficient via the Cl + $O_3$ reaction is determined to be $(1.04 \pm 0.06) \times 10^{-11}$ cm$^3$ molecule$^{-1}$ s$^{-1}$ under the present experimental conditions of 33.2 Torr and 296 K. This result agrees well with previous experimental reports [35,36] and the current JPL/NASA recommended value [37], demonstrating that microsecond-resolved dual-comb spectroscopy in the core molecular fingerprint region near 12 μm is a feasible approach for accurate kinetic studies of fast radical reactions.

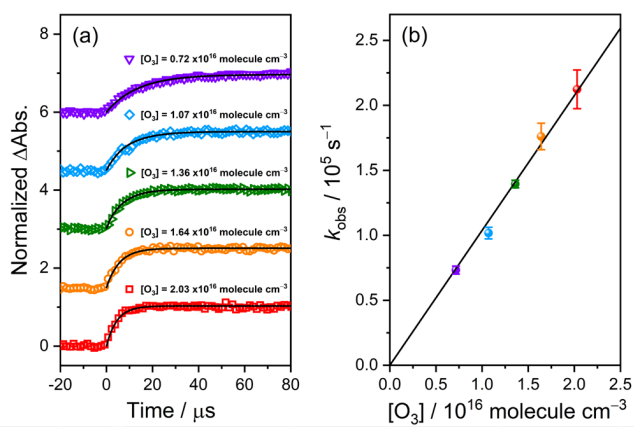

Fig. 5. (a) Temporal profiles of ClO with a temporal resolution of 1.5 μs, extracted from time-resolved dual-comb spectra recorded at a total pressure of 33.2 Torr and 296 K for different initial $O_3$ concentrations. The black curves show fits to a single-exponential rise function. (b) Observed rate coefficients ($k_{obs}$) as a function of the initial $O_3$ concentration. The black line shows the linear fit, yielding a slope of $(1.04 \pm 0.02) \times 10^{-11}$ cm$^3$ molecule$^{-1}$ s$^{-1}$.

## 3.  Conclusion

In summary, we have extended microsecond-resolved electro-optic dual-comb spectroscopy into the 10–12.5 μm molecular fingerprint region by using OP-GaP-based difference-frequency generation operated near a turning-point quasi-phase-matching condition. This approach enables broad tuning and high-resolution, time-resolved measurements in a spectral window important for reactive halogen species, yet previously difficult to access with DFG-based EO dual-comb systems. Using this dual-comb platform, we demonstrate high-resolution, time-resolved spectroscopy of transient ClO in its fundamental band near 12 μm and further apply it to kinetic measurements of rapid ClO formation in the Cl + $O_3$ reaction system by analyzing ClO time traces recorded with a temporal resolution of 1.5 μs. The agreement of the derived rate coefficient with established literature values further validates the quantitative capability of the present method for kinetic studies. Such a dual-comb platform, capable of high-spectral-resolution, microsecond-resolved measurements in the 10–12.5 μm fingerprint region, is a promising tool for precise studies of gas-phase chemical kinetics and reaction mechanisms. It also provides a route to microsecond-resolved investigations of atmospherically important halogen-oxide chemistry, including both the identification of transient halogen oxides with strong long-wavelength infrared signatures and studies of key reactions such as ClO + NO$_x$ [24,38], ClO + HO$_2$ [39], and halogen-oxide cross reactions in Earth's atmosphere, as well as related chlorine chemistry relevant to Venus's atmosphere [18].

**Funding.** National Science and Technology Council, Taiwan (grant No. 113-2628-M-001-006-MY3; 114-2923-M-001-001-MY4; 114-2927-I-001-524; 114-2639-M-A49-002-ASP) and Academia Sinica (grant No. AS-CDA-114-M05).

**Disclosures.** The authors declare no conflicts of interest.

**Data availability.** Data underlying the results presented in this paper are not publicly available at this time but may be obtained from the authors upon reasonable request.